\documentclass[12pt,preprint]{aastex}

\newcommand {\kms} {\,{\rm km\,s}^{-1}}

\usepackage{rotating}

\shorttitle{BSSs in M30}
\shortauthors{Lovisi et al.}
 
\begin{document} 
\title{FLAMES and XSHOOTER spectroscopy along the two BSS sequences of M30
\footnote{Based on FLAMES and XSHOOTER observations collected at the European Southern Observatory, proposal numbers 
087.D-0748 and 089.D-0306.}}
\author{
L. Lovisi\altaffilmark{1},
A. Mucciarelli\altaffilmark{1}, 
B. Lanzoni\altaffilmark{1},
F.R. Ferraro\altaffilmark{1},
E. Dalessandro\altaffilmark{1},
L. Monaco\altaffilmark{2}
\affil{\altaffilmark{1} Dipartimento di Fisica \& Astronomia, Universit\`a degli Studi
di Bologna, viale Berti Pichat 6/2, I--40127 Bologna, Italy}
\affil{\altaffilmark{2}European Southern Observatory, Casilla 19001, Santiago, Chile}
} 
\date{}
 
\begin{abstract} 
We present spectroscopic observations acquired with FLAMES and XSHOOTER at the Very Large Telescope 
for a sample of 15 Blue Straggler Stars (BSSs) in the globular cluster (GC) M30. 
The targets have been selected to sample the two BSS sequences discovered, 
with 7 BSSs along the blue sequence and 8 along the red one.
No difference in the kinematical properties of the two groups of BSSs has been found. In particular, 
almost all the observed BSSs have projected rotational velocity lower than $\sim$30 $\kms$, with only one (blue) 
fast rotating BSS ($>$90 $\kms$), identified as a W UMa binary. This rotational velocity distribution 
is similar to those obtained in 47 Tucanae and NGC~6397, while M4 remains the only GC 
studied so far harboring a large fraction of fast rotating BSSs.
All stars hotter than $\sim$7800 K (regardless of the parent BSS sequence) show iron abundances 
larger than those measured from normal cluster stars, with a clearcut trend with the effective temperature. 
This behaviour suggests that particle trasport mechanisms driven by radiative levitation occur in 
the photosphere of these stars, as already observed for the BSSs in NGC~6397.
Finally, 4 BSSs belonging to the red sequence (not affected by radiative levitation)
show a strong depletion of [O/Fe], with respect to the abundance measured in Red Giant Branch and Horizontal Branch stars. 
This O-depletion is compatible with the chemical signature expected in BSSs formed by mass transfer processes
in binary systems, in agreement with the mechanism proposed for the formation of BSSs in the red sequence.

\end{abstract} 
 

\section{Introduction}
\label{intro}
Blue Straggler stars (BSSs) are brighter and bluer (hotter) than the main sequence (MS) turnoff (TO).
They are located along an extension of the MS in the color-magnitude diagram (CMD) of globular clusters (GCs) and they are known to 
be more massive than normal MS stars (\citealt{Shara97, Gilliland98, DeMarco05}), thus indicating that a process able to increase 
the initial mass of a single star must be at work. While BSSs have been demonstrated to be invaluable probes of GC internal
dynamics and their radial distribution is indeed a powerful tool to measure the cluster dynamical age \citep{Ferraro2012}, 
many basic questions about the formation mechanisms and the physical properties of these puzzling objects remain open.
At the moment, two main leading scenarios have been proposed for their formation: BSSs could be generated by mass-transfer 
activity in binary systems (MT-BSSs; \citealt{McCrea64, Zinn76}), possibly up to the complete coalescence of the two companions, 
or they may form by collision-induced stellar mergers (COL-BSSs; \citealt{HillsDay76}).\\ 
The spectroscopic analysis seems to be the most promising way to discriminate 
between the two formation channels. In fact, hydrodynamical simulations predict carbon (C) and oxygen (O) depletion
for MT-BSSs, due to the fact that the material on the BSS surface comes from the deep regions of the donor star, 
where the CNO-burning already occurred (\citealt{SarnaDeGreve96}). Conversely, very little mixing is expected to occur between the inner cores 
and the outer envelopes of the colliding stars, so that COL-BSSs should show normal C and O abundances \citep{Lombardi95}.
Those theoretical results have been confirmed by the observation in 47 Tucanae \citep[47 Tuc,][hereafter F06]{Ferraro2006},
of 6 (out of 42) BSSs showing C and O depletion. Unfortunately,
at the moment, 47 Tuc is the only GC where such a kind of signature has been oberved. In fact, the subsequent analysis of the BSS population
of M4 lead to the conclusion that all of the observed BSSs show normal C and O abundances \citep[hereafter L10]{Lovisi10}, suggesting that either they 
are all formed through collisions or, more likely, that the depletion is a transient phenomen (in fact, 
according to the percentage of CO-depleted BSSs found in 47 Tuc, there should have been 0-1 BSSs with C and O depletion in M4). 
Moreover, from the analysis of the BSS population in NGC 6397,
\citet{Lovisi12} (hereafter L12) highlighted the occurrence of a particle transport process called radiative levitation 
\footnote{Please note that some authors refer to ``radiative diffusion'' or ``atomic diffusion processes'' to indicate 
the competition between radiative levitation and gravitational settling (a process that makes chemical elements on the top 
of the stellar surface to sediment on the bottom because of the gravity acceleration). In this work we distinguish the two 
processes and we refer to radiative levitation as the process occurring when the radiative acceleration overcomes the gravitational one.}
on the atmosphere of BSSs hotter than $\sim$ 8000 K. Such process, already known to occur in HB stars hotter than $\sim 11000$ K
\citep{Behr00a, Behr00b, Pace06}, alters the surface chemical abundances (in particular, it increases the mean metallicity) 
thus preventing the interpretion of any surface abundance in terms of BSS formation mechanism.\\

Another interesting property that can be studied to infer information on the BSS formation processes is the projected
rotational velocity v$_{e}$ $\sin(i)$. However, the theoretical scenario is quite complex. In fact, MT-BSSs 
are expected to have high rotational velocities because of the angular momentum transfer \citep{SarnaDeGreve96}.
Unfortunately accurate simulations are lacking, mostly because of the difficulty in following the evolution of a 
hydrodynamic system (the mass transfer between binary components) up to the system merge \citep{Sills05}.
This time-scale is unknown but certainly large, of the order of half a billion years \citep{Rahunen81}.
According to some authors \citep{BenzHills87}, also COL-BSSs should rotate fast. Nevertheless, the rotation of both MT and COL-BSSs 
might be slowed down by braking mechanisms (\citealt{LeonardLivio95, Sills05}),
like magnetic braking and disk locking, for which efficiencies and time-scales are not well known yet.

From an observational point of view, a large fraction of fast rotating BSSs ($\sim$ 40\%) has been identified in M4 by L10
but, at the moment, this is the only studied cluster exhibiting such a large population of fast rotating BSSs. In fact, according to
F06 and L10, almost all the BSSs in 47 Tuc and NCG 6397 have low rotational velocities, with only one
exception. Moreover, no difference has been found in the rotational velocity distributions of CO-depleted and normal BSSs
in 47 Tuc, suggesting that no particular rotational velocity can be univocally linked to a given formation channel.\\

However, the discovery of two distinct sequences of BSSs in the GC M30 \citep[][hereafter F09]{Ferraro09} has recently opened
a new perspective in the study of a possible photometric signature of the BSS formation process:
two almost parallel sequences similarly populated (24 stars in the blue sequence and 21 stars in the red one)
have been revealed in this cluster. The two populations show a different level of segregation (the red BSSs
being more centrally concentrated) and possibly have different origin since the blue sequence is well fit by
evolutionary models of COL-BSSs (\citealt{Sills09}), while the red one well corresponds to the lower
envelope of the distribution of MT binaries simulated by \citet{Tian06}. F09 suggested that
the core collapse event boosted the rate of direct stellar collisions, hardened binary systems
and could have generated the two populations 1-2 Gyr ago.

The present study is devoted to a spectroscopic screening of BSSs selected along the two sequences,
with the goal of investigating their physical properties and further investigate the proposed interpretation
of the two distinct populations.
The paper is organized as follow: the observations are described in Sect. \ref{obs}. The
determination of stellar radial velocities and cluster membership are
discussed in Sect. \ref{rad}. The estimate of atmospheric
parameters to all targets is described in Sect. \ref{par}. The
measured rotational velocities are presented in Sect. \ref{rot}.
Sect. \ref{chem} describes methods and results of the chemical abundance analysis for our
sample. Finally our conclusions are drawn in Sect. \ref{discuss}.

\section{Observations}
\label{obs}

\begin{figure}
\plotone{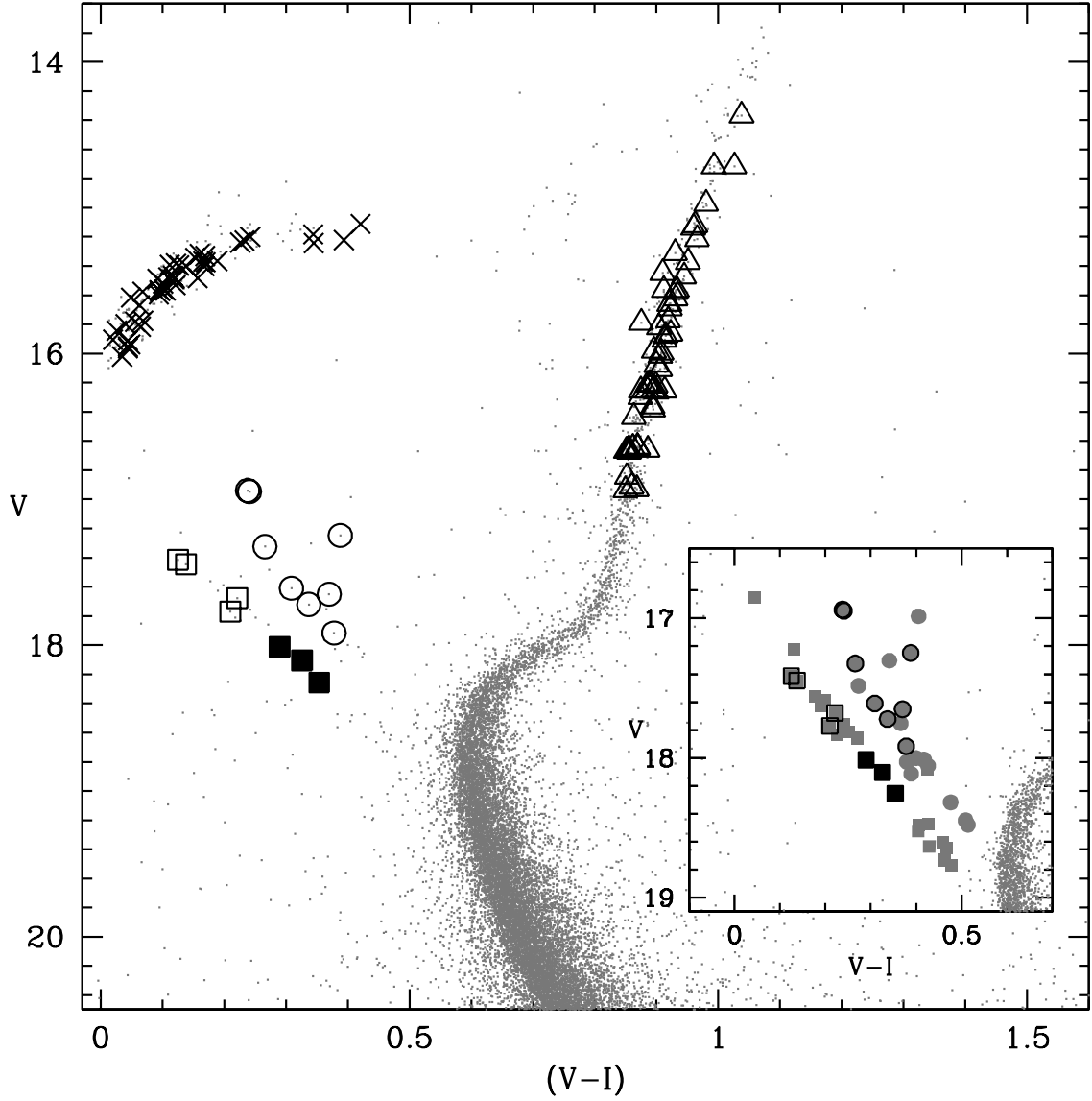}
\caption{Colour-magnitude diagram of M30. Squares mark the observed BSSs in the blue sequence whereas
circles are BSSs in the red one. Black squares are the XSHOOTER targets. The observed RGB stars are depicted with triangles and the HB ones with crosses.
The inset shows a zoom on the BSS region.}
\label{cmdm30}
\end{figure}

This work is based on spectroscopic observations collected with the spectrographs 
FLAMES and XSHOOTER both mounted at the Unit Telescope 2 at the VLT of the European 
Southern Observatory.
The spectroscopic targets have been selected from a photometric catalog obtained by
combining WFPC2@HST data for the central cluster region and MegaCam (at the Canada-France-Hawaii 
 telescope) data for the outer region (see F09 for details).
\begin{enumerate}
\item {\sl FLAMES dataset ---}
The observations have been performed with the multi-object high-resolution 
spectrograph FLAMES in the UVES+GIRAFFE combined mode, during the nights 7-8-9 August 2011. 
The FLAMES fibres allocation has been made to maximize
the number of observed BSSs in both the blue and the red sequence.
Unfortunately, 
the majority of the BSSs are concentrated in the inner 30\arcsec\ (F09) and some
of them have close bright companions. Moreover, the physical size of the FLAMES magnetic buttons 
prevents us to simultaneously observe targets that are closer than 11$\arcsec$.
We conservatively excluded targets having stellar sources of comparable or brighter
luminosity within 3 arcsec.
The FLAMES sample includes 12 BSSs, 52 Red Giant Branch (RGB) and 49 Horizontal Branch (HB) stars. Among the observed BSSs, 
4 targets belong to the blue sequence and 8 to the red one. 
Fig. \ref{cmdm30} shows the CMD of M30 with the selected targets highlighted.

Two different setups have been used for the spectroscopic observations:
HR5A (with spectral coverage 4340-4587 \AA\ and spectral resolution R$\simeq$18500) and HR18 (7468-7889 \AA , R=18400), 
suitable to sample some metallic lines and the O~I triplet at $\lambda \simeq 7774$ \AA ,
respectively. Exposure times amount to 4.5 hours for the HR5A setup, and to 3 hours
for the HR18. The spectra pre-reduction has been done by using
the latest version of the GIRAFFE ESO pipeline, gasgano version 2.4.3, \footnote{http://www.eso.org/sci/software/pipelines/} 
that includes bias subtraction,
flat-field correction, wavelength calibration and one-dimensional spectra extraction. 
For each exposure we subtracted the corresponding master sky spectrum, obtained by averaging 
several spectra of sky regions.
By combining the sky-subtracted exposures, we
obtained median spectra with signal-to-noise ratios S/N$\simeq 30-70$
for the selected BSSs, and S/N$\gtrsim 80$ for the RGB and HB stars.

\item {\sl XSHOOTER dataset ---}
Observations of three BSSs belonging to the blue sequence have been secured during 
the nights 27 and 28 June 2012 with XSHOOTER. Those targets are too faint (V$\sim$18.0--18.3) to be efficiently observed with FLAMES 
in a reasonable exposure time, but some information can be inferred from XSHOOTER spectra, in spite of the lower 
resolution of this spectrograph. The three targets have been observed in stare mode enabling 
simultaneously the UVB ($\sim$3300-5500 \AA) and the VIS ($\sim$5500-10000 \AA) channels. 
The slit width was set at 0.8$\arcsec$ (R=~6200) and 0.7$\arcsec$ (R=~11000) for the UVB and the VIS channels, respectively.
The exposure times range from 45 to 90 minutes, according to the magnitude of the target and 
the seeing conditions.
The spectral reduction has been performed with the version 2.0.0 of the XSHOOTER ESO pipeline 
\citep{modigliani}, including bias-subtraction, flat-field correction, wavelenght calibration, 
correction for sky-background and order-merging.
The spectral extraction has been performed manually with the IRAF task {\sl apall}. 
In Fig. \ref{cmdm30} the three BSSs observed with XSHOOTER are marked with black squares.
\end{enumerate}

\section{Radial velocities and cluster membership}
\label{rad}
In order to assess the cluster membership for each target, we measured the radial velocities (RVs)
with the IRAF task \textit{fxcor}. Synthetic spectra calculated with the atmospheric parameters of the 
analysed targets (see Sect. \ref{par}) and convolved with a Gaussian profile to reproduce the spectral resolution 
of the observed spectra, have been used as templates for the cross-correlation (see Sect. \ref{rot} for details about 
the computation of the synthetic spectra).
Finally, the heliocentric correction has been applied to all the radial velocities. 

\begin{figure}
\plotone{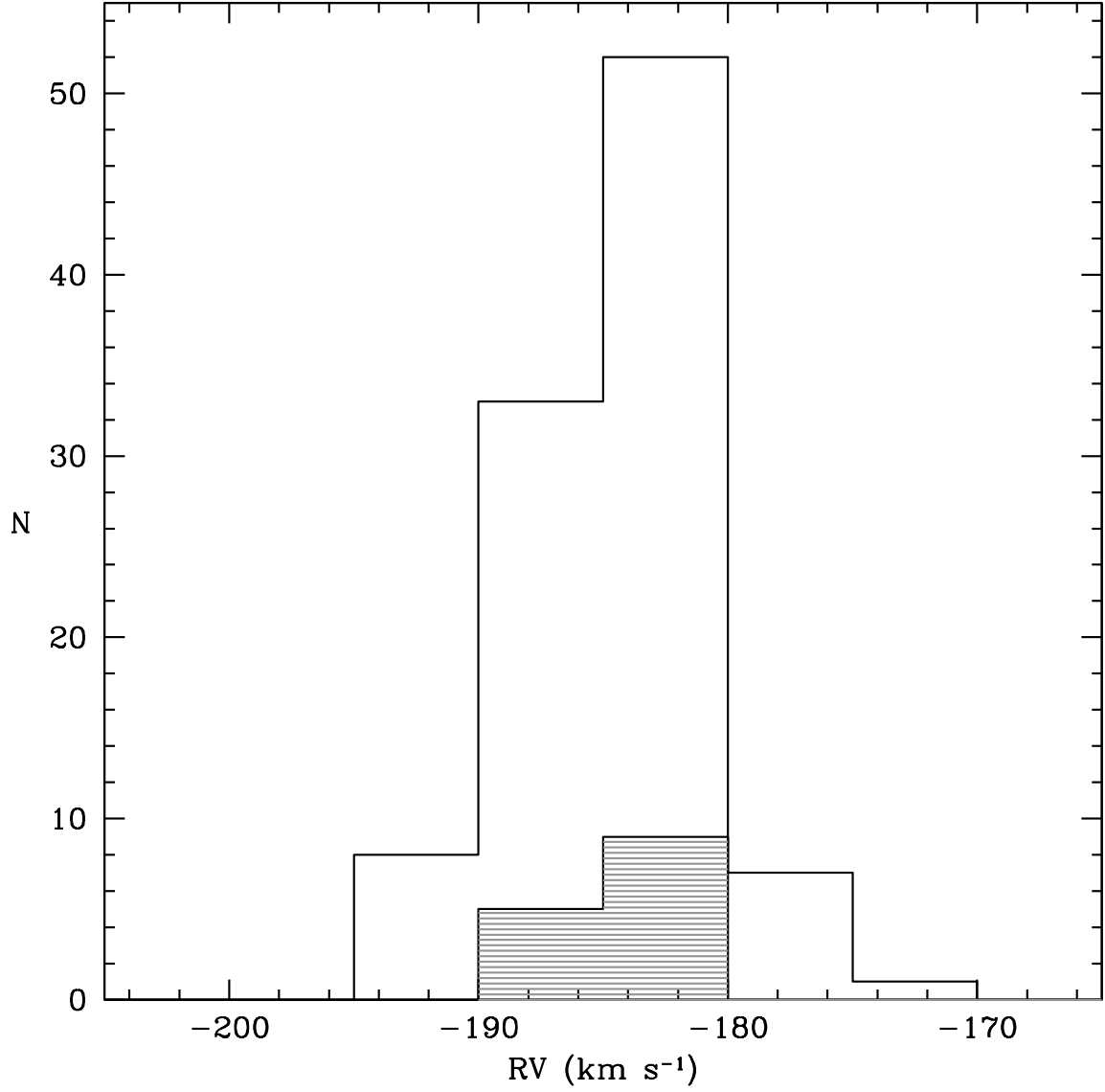}
\caption{Radial velocity distributions for the RGB+HB stars (empty histogram) and BSSs (grey shaded histogram).} 
\label{vradm30}
\end{figure}

The derived RV distribution is shown in Fig. \ref{vradm30}. 
The mean radial velocity of the RGB stars is RV$=-$184.9$\pm$0.5 $\kms$ ($\sigma$=3.4 $\kms$)  
which is fully in agreement with that obtained for the HB stars (RV$=-$184.6$\pm$0.5 $\kms$, $\sigma$=3.3 $\kms$).
The value for the combined RGB+HB star sample is RV$=-$184.7$\pm$0.3 $\kms$ ($\sigma$=3.4 $\kms$
and it is in agreement with previous results by \citet{Harris96}, \citet{Zaggia92} and \citet{Carretta09a, Carretta2009b}.
This value has been assumed as the systemic velocity of M30 and used to infer the cluster membership
for each star: stars having RVs within 3$\sigma$ with respect to the systemic velocity value have been considered as members of M30.
The BSS RV distribution is in good agreement with that of
the RGB+HB stars, with a mean value RV$=-$184.4$\pm$0.7 $\kms$ ($\sigma$=2.7 $\kms$). 
All the observed BSSs turned out to be cluster members. Also, the RVs of the BSSs 
in the two sequences  nicely agree with each other: 
we measured RV$=-$184.6$\pm$1.3 $\kms$ ($\sigma$=3.11 $\kms$) for blue BSSs and 
RV$=-$184.2$\pm$0.9 $\kms$ ($\sigma$=2.5 $\kms$) for the red ones.

BSS \#12005407 deserves a particular discussion. FLAMES spectra of this star appear featureless 
and we are not able to derive its RV (and not to perform the chemical analysis). 
The observed counts of its spectra are fully compatible with those of the other BSSs, 
thus excluding that the lack of spectral features is due to a wrong centering of the fiber.
This star has been classified as W Uma variable by \citet{Pietrukowicz04}. 
Hence, the complete lack of lines in all the observed setups is likely due to a very high rotational 
velocity (which is indeed commonly associated to this class of objects).
In spite of this, on the basis of its central position (the BSS is located at $\simeq$ 75\arcsec\ from 
the cluster centre), it has been considered a cluster member.

\section{Atmospheric parameters}
\label{par}
Effective temperature (T$_{eff}$) and surface gravity (log g) for all the observed targets have been derived 
comparing their position in the CMD with theoretical isochrones (with different ages) and zero age horizontal branch (ZAHB) 
models from the Pisa Evolutionary Library \citep{Cariulo04} \footnote{This dataset of theoretical isochrones is the one adopted by F09 to fit the evolutionary 
sequences in the CMD of M30.}. 
Given the stellar position in the CMD, each observed target has been orthogonally projected on the closest 
theoretical model, and T$_{eff}$ and log g have been derived for each star. 
For the RGB stars, an isochrone of 12 Gyr, Z=0.0002 (compatible with the measured [Fe/H]=$-$2.3 for the cluster, \citealt{Carretta09a})
and $\alpha$-enhanced chemical mixture has been superimposed on the CMD of M30, assuming a distance modulus 
(M-m)$_{0}=14.71$ and E(B-V)=0.03 \citep{Ferraro1999}. The same distance modulus and reddening
have been applied to a ZAHB model with the same metallicity and chemical mixture to derive
atmospheric parameters for the HB stars. For BSSs, isochrones with different ages (and assuming the same 
distance modulus and reddening values) have been used to sample the BSS region.

The atmospheric parameters for the BSS, RGB and HB samples are listed in Tables \ref{bssm30}, \ref{rgbm30} 
and \ref{hbm30}, respectively.
Conservative errors in temperatures have been considered: 100 K for BSSs and RGB stars and 50 K for HB stars.
Errors in gravities are negligible for all targets.
Concerning microturbulent velocity, the small number of metallic lines (due to the low metallicity of the cluster) 
prevents us to derive spectroscopically this parameter. Microturbulent velocities for the RGB stars have been derived from the 
relation presented by \citet{Kirby09}, expressing the microturbulent velocity as a function of the surface gravity.
For BSSs and HB stars it has been assumed equal to $2 \kms$. However the assumption of a different 
value of microturbulent velocity has a negligible impact on rotational velocities and chemical abundances 
and it does not change our results. We finally adopted a conservative error of 0.5 $\kms$ for all the targets. 

\section{Rotational velocities} 
\label{rot}
Rotational velocities have been derived for all the FLAMES targets in our sample but BSS \#12005407,
for which only an upper limit has been obtained. Conversely, only lower limits have been derived for the XSHOOTER targets,
due to the lower resolution of the spectra that prevents us to appreciate small differences at low rotational velocities. 

Rotational velocities for the RGB stars have been inferred from the Ba~II at 4554.03 \AA\ whereas 
rotational values for the BSSs from some strong metallic lines, such as the Ti~II lines at 4501.270,
4468.492 and 4571.971 \AA\ and the Mg~II line at 4481 \AA.
The latter Mg~II line has been used also to infer rotational velocities for the HB stars. 
For each star, we computed a grid of synthetic spectra
with suitable atmospheric parameters and different rotational velocities. A $\chi^{2}$ minimization between 
the observed spectra and the synthetic ones is then performed and used to determine the best-fit rotational velocity. 

All the synthetic spectra have been computed by using the Kurucz's code SYNTHE \citep{kurucz93, Sbordone04}, 
including the atomic data for all the lines from the most updated version of the Kurucz line list by F.Castelli 
\footnote{http://wwwuser.oat.ts.astro.it/castelli/linelists.html}. 
Model atmospheres for the synthetic spectra have been computed with ATLAS9 code \citep{kurucz93, Sbordone04} under the assumptions of Local Thermodynamic Equilibrium 
(LTE) and plane-parallel geometry and adopting the new opacity distribution functions by \citet{CastelliKurucz2003}, without the 
inclusion of the approximate overshooting \citep{Castelli1997}. Typical uncertainties in the v$_{e}$ $\sin(i)$ measurement 
are of the order of 1-2 $\kms$.

\begin{figure}
\plotone{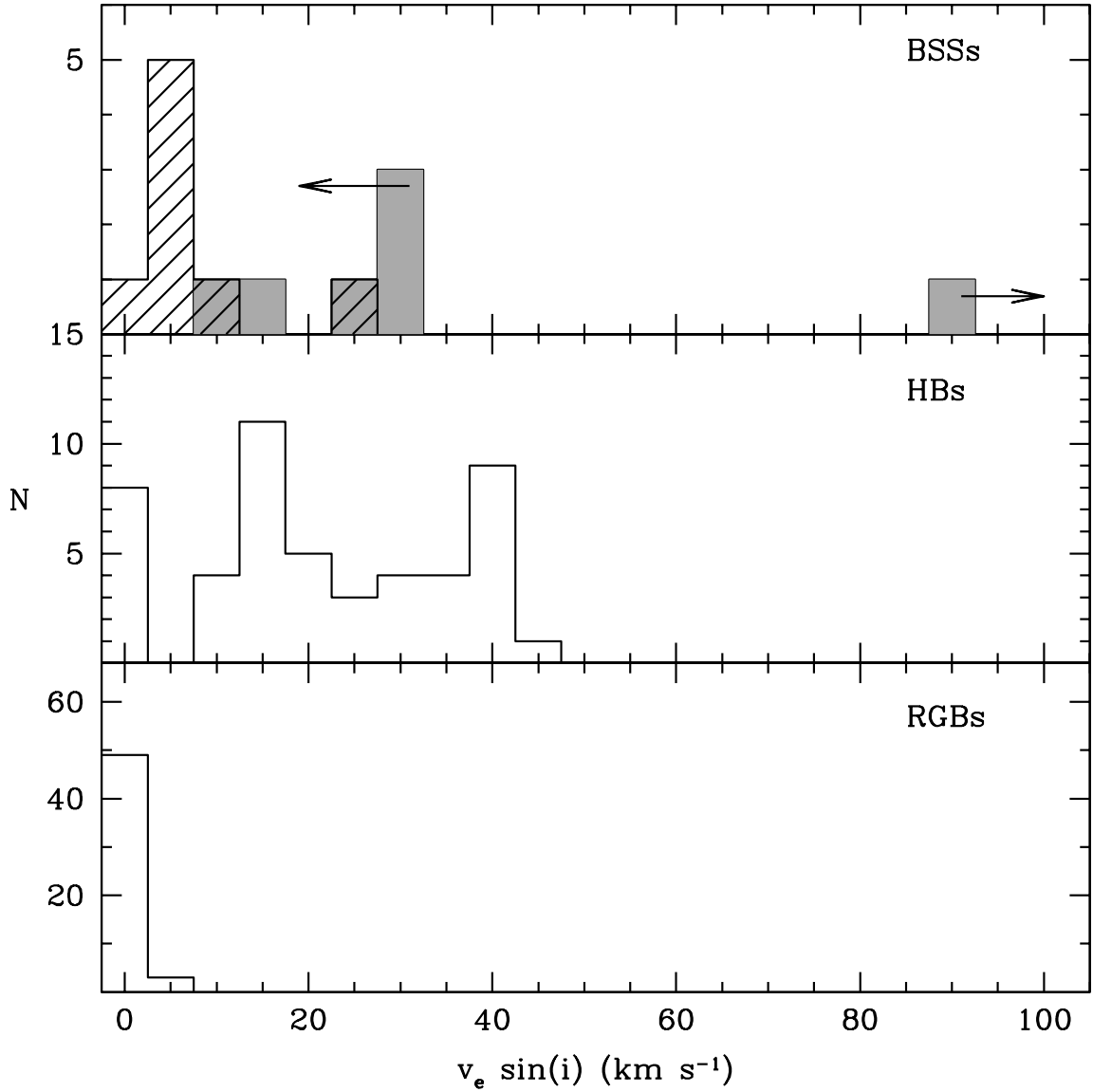}
\caption{Rotational velocity distributions for BSSs in the blue and the red sequence (upper panel, 
filled and shaded histogram, respectively), HB stars (central panel) and RGB stars (lower panel) of our sample.
Arrows mark lower or upper limits.} 
\label{rotm30}
\end{figure}

The rotational velocity distribution thus obtained for BSSs is shown in Fig. \ref{rotm30} (upper panel)
and it is compared with that of HB stars (central panel) and RGB stars (lower panel). 
Almost all RGB stars have v$_{e}$ $\sin(i) = 0 \kms$, the largest value not exceeding 7 $\kms$. 
The distribution for the HB stars is wide, ranging between $0 \kms$ and $47 \kms$. 
These results are in agreement with those usually observed for this kind of stars
\citep[L12]{Peterson95, Behr00a, Behr00b, Carney08, Cortes09, Mucciarelli11}. 

Also the rotational velocity distribution for BSSs is spread with most red and blue BSSs (shaded and grey histograms,
respectively) having rotational velocity between 0 and 30 $\kms$ with the only exception of the fast rotating blue BSS (\#12005407): 
for this star, the upper limit suggests a rotational velocity larger than 90 $\kms$. 

Finally, Fig. \ref{fem30} shows the rotational velocity as a function of temperature for all our targets
and reveals no clear trends.

\section{Chemical abundances}
\label{chem}
For all the FLAMES targets we derived Fe, Mg, Ti and O abundances or upper limits. 
Abundances of Fe, O and Ti for all the targets as well as Mg abundances for the RGB stars have been estimated 
with the code GALA \citep{Mucciarelli2013} by using the measured equivalent widths (EWs) of the absorption lines.
Conversely, spectral synthesis has been used to derive the Mg abundance of BSSs and HB stars 
from the Mg~II line at $\sim$ 4481 \AA, that is an unresolved blend of three close components 
of the same multiplet (so that the observed line profile significantly deviates from the Gaussian approximation ).
A $\chi^{2}$ minimization has been performed by using a grid of synthetic spectra computed with the appropriate 
atmospheric parameters and rotational velocities.
Finally for the oxygen of the BSS FLAMES targets, we were able to obtain only upper limits, 
corresponding to a 3$\sigma$ detection according to the formula by \citet{Cayrel88}. 

For the three XSHOOTER targets, the combination of low metallicity and low spectral resolution does allow to analyse only a few transitions. 
We measured 2-3 Fe~II lines and the Mg~II line at $\sim$ 4481 \AA. For the coldest BSS (namely \#13000594) 
we derived only upper limits for both Fe and Mg.

The atomic parameters of the analysed lines are listed in Table \ref{bsslines}, \ref{rgblines} and \ref{hblines}.
The used reference solar abundances are from \citet{GrevesseSauval1998} for Fe, Mg and Ti, and from 
\citet{Caffau2011} for O.

An important issue to take into account is the possible deviation from the LTE assumption, particularly for the 
hottest targets. For the O limits we included non-LTE corrections taken from the statistical equilibrium calculations of \citet{Takeda1997}. 
For Fe, Mg and Ti abundances, no grid of non-LTE corrections are available in the literature for the range of parameters typical of our targets.
Based on the fact that non-LTE corrections for ionized elements in hot (A and F type) stars are negligible with respect to those for
neutral elements, whereas the opposite is true for cold stars, we obtained Fe and Mg abundances for the BSSs and HB stars 
by using Fe~II and Mg~II lines, whereas Fe~I and Mg~I lines have been used for the RGB stars. 
We did the same for Ti abundances in BSSs and HB stars whereas, unfortunately, this was
not possible for RGB stars since only lines from ionized elements transitions are present in the spectra. The derived 
abundances are listed in Tables \ref{bssm30}, \ref{rgbm30} and \ref{hbm30} for BSSs, RGB and HB stars, respectively. 
 
\begin{figure}
\plotone{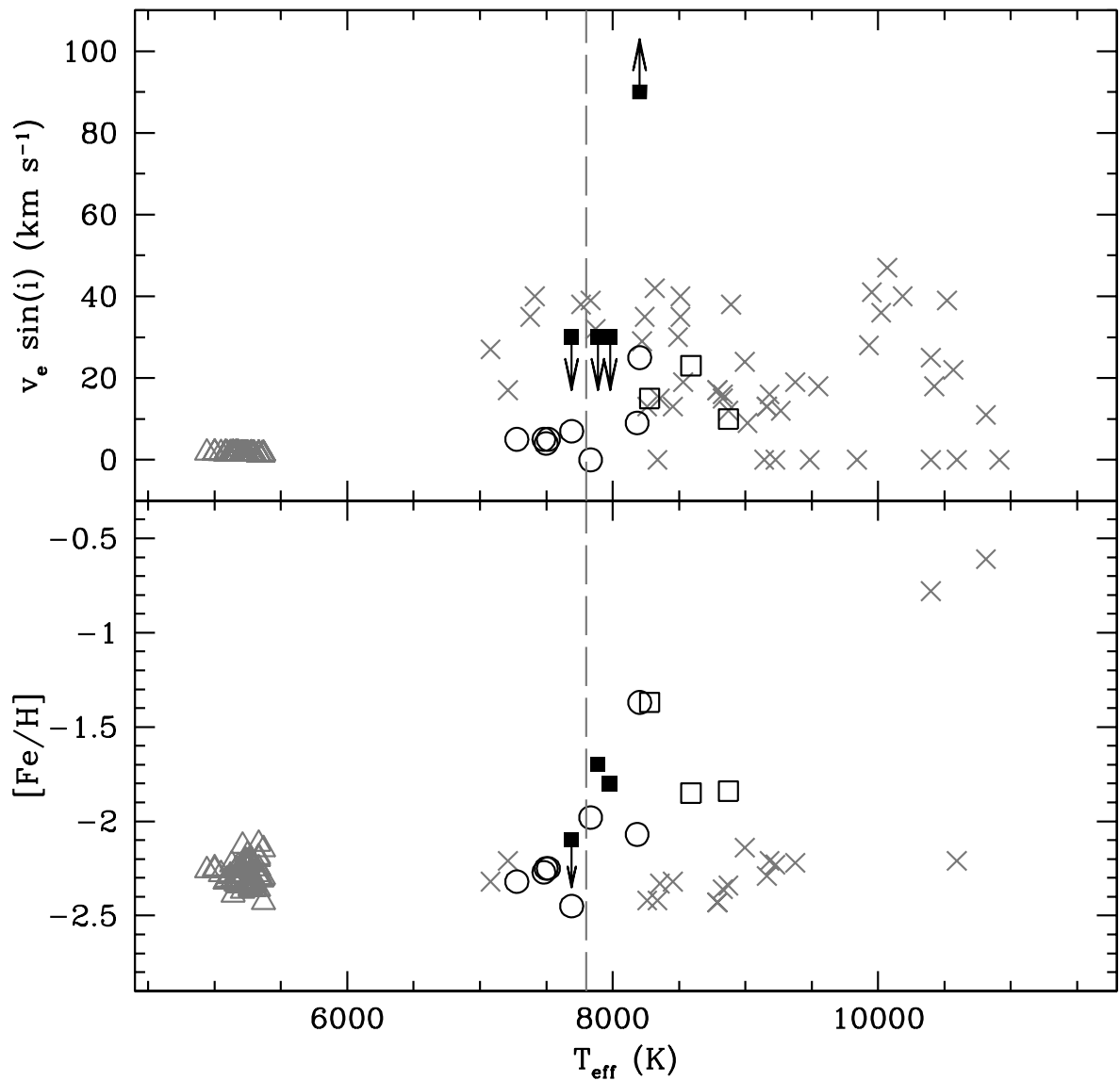}
\caption{Upper panel: rotational velocities as a function of the effective temperature for RGB stars 
(grey triangles), HB stars (grey crosses) and BSSs in the two sequences 
(squares and circles for the blue and red BSSs, respectively). Lower panel: [Fe/H] ratio 
as a function of T$_{eff}$ for the same targets. The dashed line mark the observed limit for the occurence of the radiative levitation.} 
\label{fem30}
\end{figure}

{\bf Iron --}
For the RGB stars we find an average iron abundance of [Fe/H]=$-$2.28$\pm$0.01 ($\sigma$=0.07) dex in very good agreement with 
results by \citet{Carretta09a, Carretta2009b, Carretta2009c}. The mean value for the HB stars is slightly larger, corresponding to 
[Fe/H]=$-$2.13$\pm$0.12 ($\sigma$=0.53). However, the large dispersion of the measures is essentially due to two out of the 
three stars with T$_{eff}>$10000 K, (see Fig. \ref{fem30}, bottom panel), suggesting that radiative levitation is affecting the 
surface abundances of those stars. In fact, by excluding the two HB stars with the largest value of [Fe/H] 
(namely \#30016296 and \#30018857), we obtain [Fe/H]=$-$2.31$\pm$0.02 ($\sigma$=0.09), in agreement with RGB stars.

Also the distribution for the BSSs is very large and the values of [Fe/H] systematically increase for incresing temperatures (Fig. \ref{fem30},
bottom panel) due to the effect of the radiative levitation. The 5 coldest (red) BSSs (T$_{eff}$ $<$ 7800 K) have 
[Fe/H]=$-$2.31$\pm$0.04 ($\sigma$=0.08), in very good agreement with RGB stars and HB stars not affected by radiative levitation. 
The same iron content is confirmed also by the upper limit found for the coldest (blue) XSHOOTER target ([Fe/H]$<-2.1$ dex).
Conversely, the hottest red BSSs (T$_{eff}$ $>$ 7800 K) have [Fe/H]=$-$1.81$\pm$0.22 ($\sigma$=0.38), 
significantly larger than the mean cluster metallicity, and the BSSs in the blue sequence have [Fe/H]=$-$1.96$\pm$0.10 ($\sigma$=0.35). 

\begin{figure}
\plotone{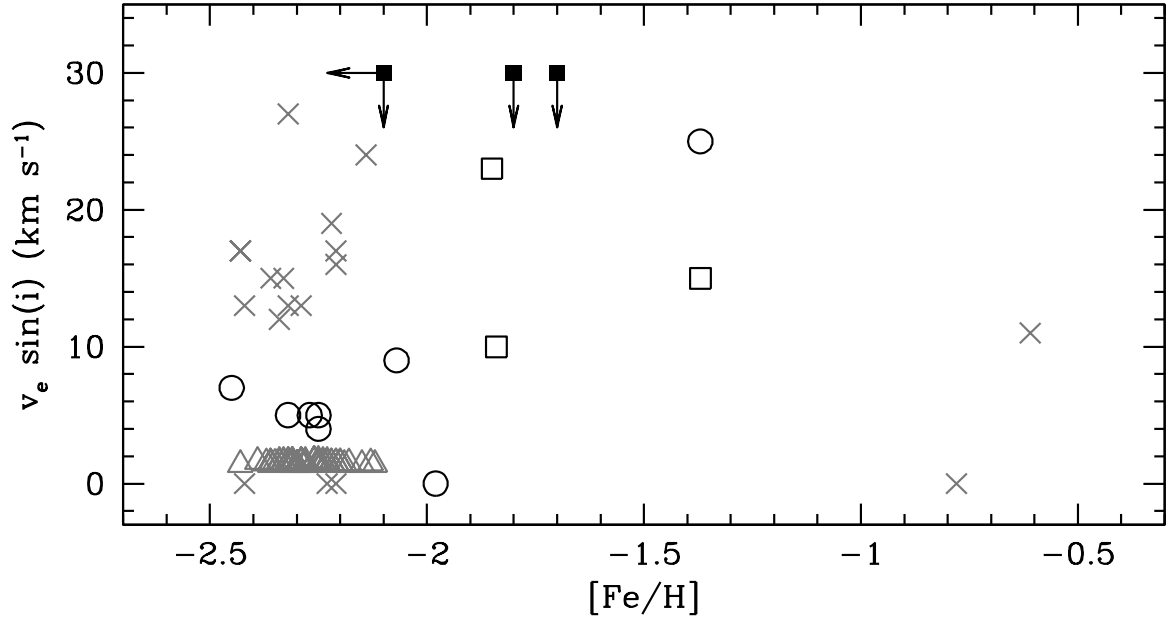}
\caption{Rotational velocity for RGB, HB stars and BSSs in two sequences as a function of the [Fe/H] ratio. Symbols are as in Fig. \ref{fem30}.} 
\label{ferotm30}
\end{figure}

In Fig. \ref{ferotm30} rotational velocities as a function of the iron content are shown. No trend between v$_{e}$ $\sin(i)$ and [Fe/H] exists
for the RGB stars. For HB stars, a mean value of $\sim$ 13 $\kms$ (with a large scatter, however) is observed for stars with [Fe/H]$<-$2.0 dex
whereas for the two more metallic stars, v$_{e}$ $\sin(i)$ drops toward values lower than $\sim$ 10 $\kms$.
Finally, a possible increase of v$_{e}$ $\sin(i)$ with increasing metallicity is observed for the BSSs.

\begin{figure}
\plotone{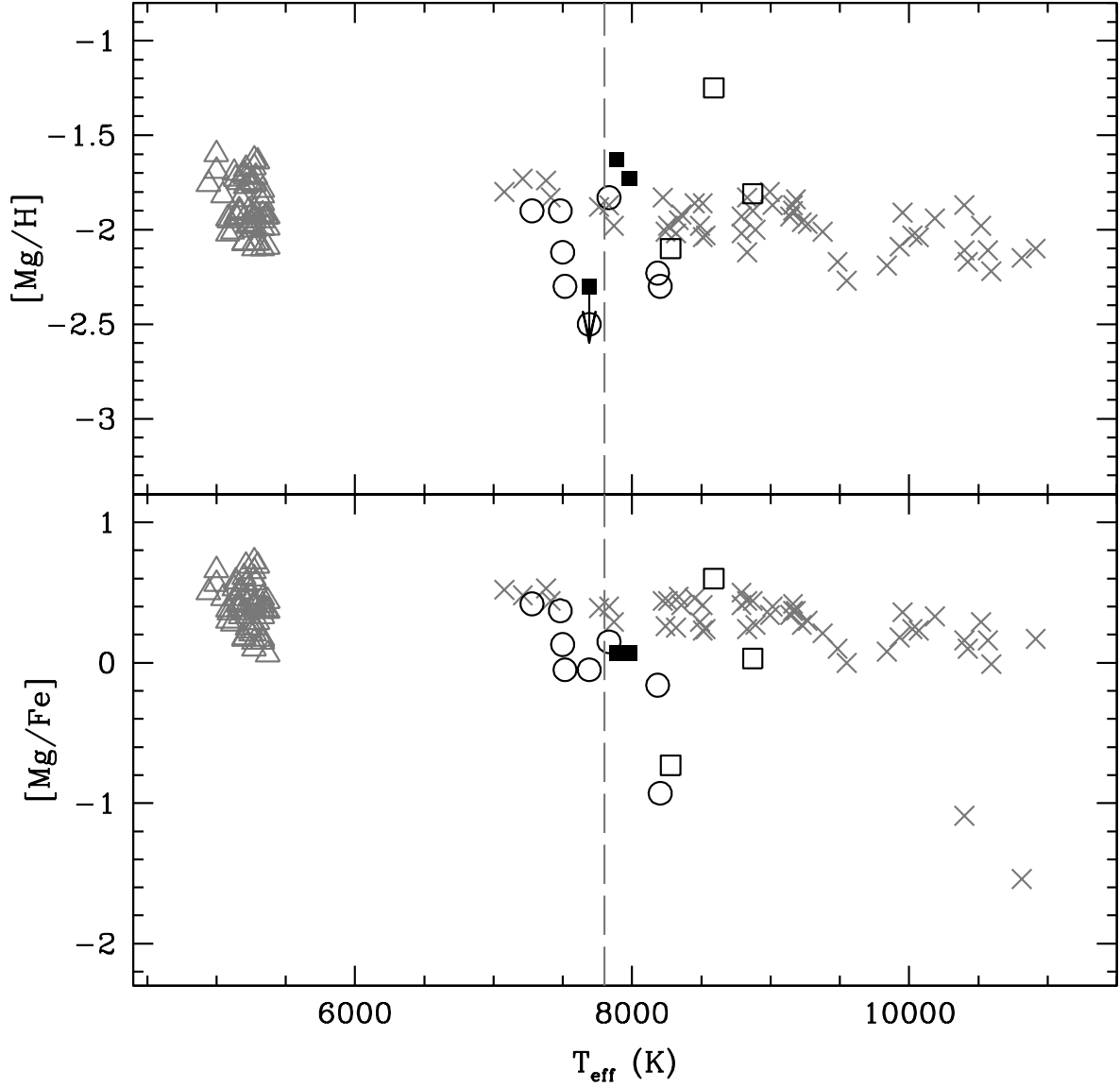}
\caption{[Mg/H] and [Mg/Fe] ratios as a function of the effective temperature for our targets (same symbols as in Fig. \ref{fem30}).} 
\label{mgm30}
\end{figure}

{\bf Magnesium --}
RGB stars have [Mg/Fe]=0.40$\pm$0.02 ($\sigma$=0.16), in good agreement with the results by \citet{Carretta2009b}. 
For all the HB stars [Mg/Fe]=0.24$\pm$0.05 ($\sigma$=0.35), whereas excluding the two HB stars affected by radiative levitation
we obtain [Mg/Fe]=0.31$\pm$0.02 ($\sigma$=0.13), in agreement with RGB stars.
For the cold red BSSs, the mean [Mg/Fe] ratio is 0.16$\pm$0.10 ($\sigma$=0.22), slightly lower than the RGB and HB values.
Significant difference in the [Mg/Fe] ratio has been found between the red cold BSSs and the red hot BSSs (affected by radiative
levitation) that show [Mg/Fe]=$-$0.31$\pm$0.32 ($\sigma$=0.56). Interestingly, that difference exclusively depends on the Fe abundances:
in fact, [Mg/H] ratio is $-$2.14$\pm$0.11($\sigma$=0.26) for the cold red BSSs and $-$2.12$\pm$0.15($\sigma$=0.25) for the hot red BSSs,
suggesting that the Mg abundances does not change with the occurrence of the radiative levitation.
Fig. \ref{mgm30} shows [Mg/H] and [Mg/Fe] ratios versus effective temperature for the RGB, HB stars and BSSs.

{\bf Oxygen --}
For BSSs, only upper limits to the O abundance (from the O~I triplet at $\sim$ 7774 \AA) can be determined because no line is 
detectable in these spectra. Also in our RGB spectra, the O~I triplet lines are too weak to be detected, thus we measured O 
abundances only for HB stars.

\begin{figure}
\plotone{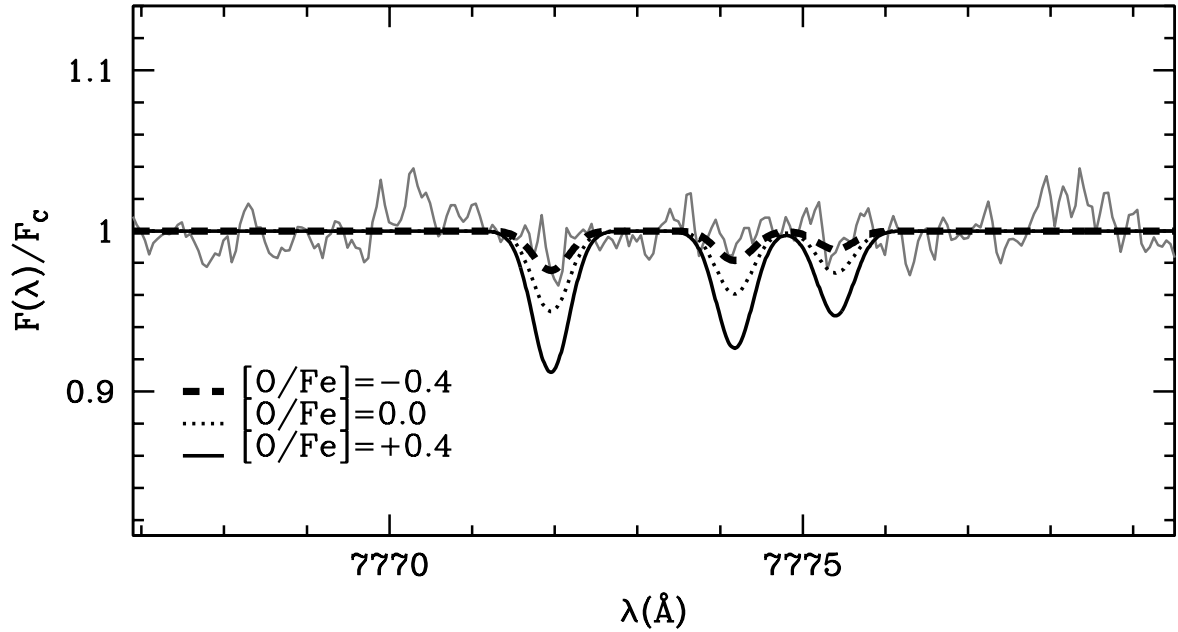}
\caption{Spectrum of BSS \#11002171 zoomed in the O~I triplet region (grey line). The dashed, dotted and solid 
lines are synthetic spectra computed with [O/Fe]=-0.4, 0.0, +0.4, respectively.} 
\label{Olines}
\end{figure}

For the HB stars (not affected by radiative levitation) we measured [O/H]=$-1.80 \pm 0.04$ ($\sigma$=0.23),
in agreement with values obtained for RGB stars by \citet{Carretta2009c}. 
As shown in Fig. \ref{oxm30} the upper limits for almost all BSSs are incompatible with the abundances of HB stars,
with the only exception of BSS \#11000416. Fig. \ref{Olines} shows the spectrum of BSS \#11002171 zoomed in 
the region of the O~I triplet. Synthetic spectra computed with atmospheric parameters and rotational velocity 
of the observed star and [O/Fe]= +0.4, 0.0, $-$0.4 dex (solid, dotted and dashed lines, respectively) have been superimposed
and clearly show that the O abundance cannot be larger than $-$0.4 dex.

\begin{figure}
\plotone{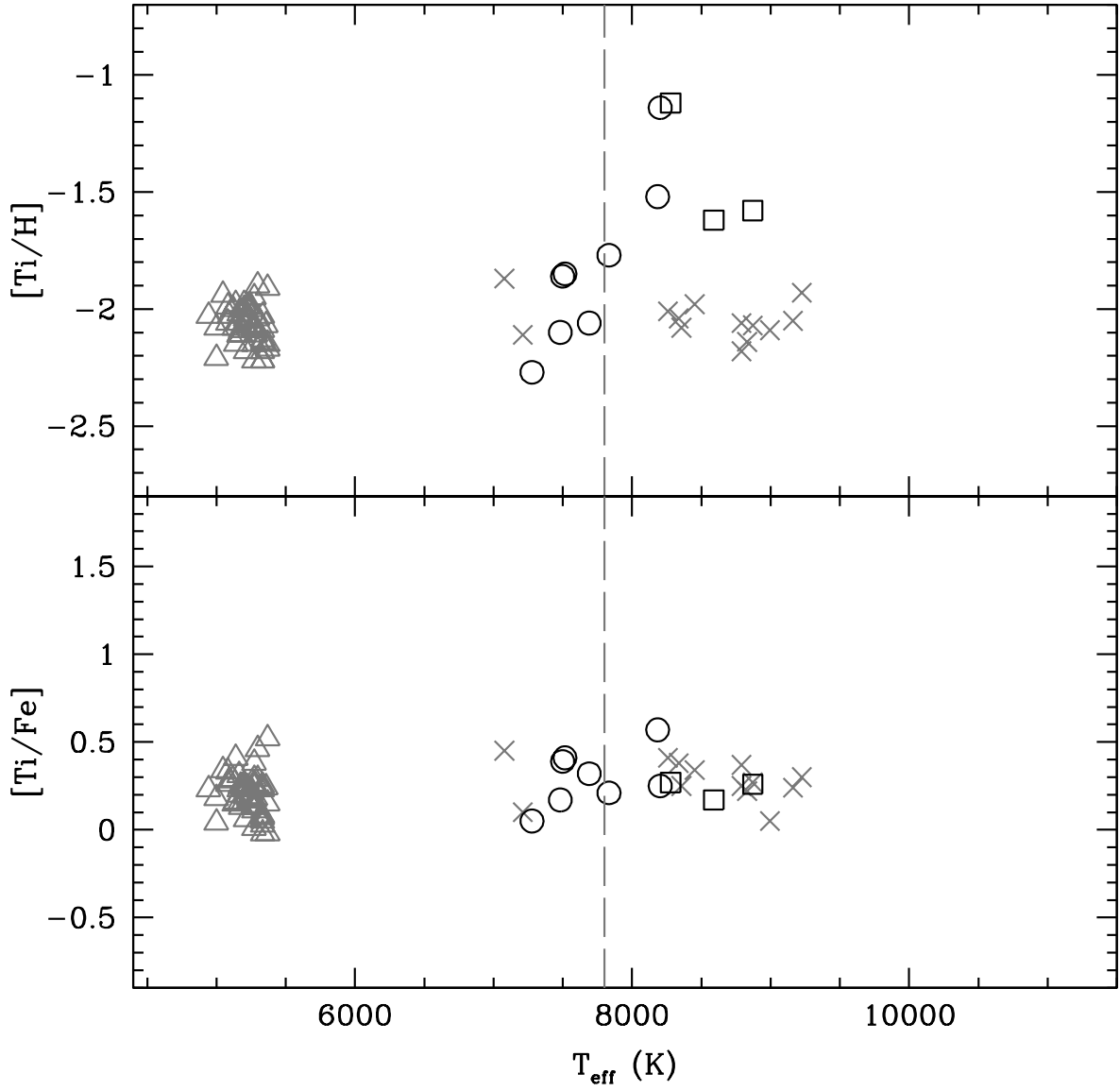}
\caption{Ti abundances for our targets (same symbols as in previous figures).} 
\label{tim30}
\end{figure}

{\bf Titanium --}
Ti abundances have been derived for all our targets by measuring the EWs of a tens of Ti~II lines.
Results are shown in Fig. \ref{tim30}. The Ti mean abundance for the RGB stars is [Ti/H]=$-$2.07$\pm$0.01 ($\sigma$=0.08), 
corresponding to [Ti/Fe]=0.20$\pm$0.01 ($\sigma$=0.11). For the HB stars [Ti/H]=$-$2.05$\pm$0.02 ($\sigma$=0.08)
and [Ti/Fe]=0.23$\pm$0.03 ($\sigma$=0.11), in agreement with RGB values.
For the cold red BSSs we derive [Ti/Fe]=0.27$\pm$0.07 ($\sigma$=0.15) corresponding to [Ti/H]=$-$2.03$\pm$0.08 ($\sigma$=0.18) 
in good agreement with the RGB and HB values. Conversely, the abundances for the hot red BSSs are larger and more scattered,
corresponding to [Ti/Fe]=0.34$\pm$0.11 ($\sigma$=0.20) and [Ti/H]=$-$1.48$\pm$0.18 ($\sigma$=0.32).
Moreover, a clearcut trend with T$_{eff}$ does exist, even if the two hottest (blue) BSSs do not have the largest Ti abundance.

\section{Discussion}
\label{discuss}
This work presents the first investigation of the kinematical and chemical properties of BSSs belonging 
to the two BSS sequences of M30. All the observed BSSs turn out to be members of the cluster, with no 
significant differences in the RV mean value and dispersion between the two groups of stars.

{\sl Rotational velocities---}
The BSSs of M30 show a rotational velocity distribution wider than that derived for the RGB stars
and more similiar to the distribution obtained for the HB stars. Most of the BSSs have  
v$_{e}$ $\sin(i) <$30 $\kms$, with the only exception of the blue BSS \#12005407 with v$_{e}$ $\sin(i) >$ 90 $\kms$ 
and classified as a W UMa. 
The rotational velocity distribution of BSSs in M30 
is very similar to that derived for the BSS populations in 47 Tuc (F06) and  NGC~6397  (L12),  where the majority of 
BSSs rotate slowly. On the contrary, the v$_{e}$ $\sin(i)$ distribution of BSSs in M4 discussed by L10 is very broad, including a large 
($\sim$40\%) fraction of {\sl fast rotators} (i.e. BSSs with v$_{e}$ $\sin(i)$ larger than $\sim$50 $\kms$). 

It is interesting to note that although the BSSs in both sequences cannot be defined fast rotators (following the
definition by L10), when we consider the rotational velocity distribution of the BSSs in each sequence, 
we find slightly higher values among the blue BSSs.
In the framework proposed by F09, the two BSS sequence of M30 are linked to different formation channels, 
the red sequence being populated by MT-BSSs and the blue sequence by COL-BSSs. 
The results obtained in this work indicate that no significant difference in rotational velocity characterize
the BSSs formed through the two mechanisms. 
The current theoretical models provide conflicting predictions about BSS rotational velocity 
\citep[see for instance][]{LeonardLivio95,Sills05}. As discussed in L12, 
high rotational velocities could be a transient stage in the life of some BSSs, that will be slowed down 
during their subsequent evolution by some braking mechanisms 
\citep[that are still poorly understood, see][]{LeonardLivio95}.

{\sl Chemical composition---}
Two main results have been derived from the chemical analysis of BSSs:\\ 
(1)~the iron abundance increases with effective temperature temperature: the 6 coldest BSSs have [Fe/H] ratios 
fully compatible with the iron content of the RGB and HB stars, while the BSSs hotter than $\sim$7800 K show 
higher iron abundances.
This trend nicely agrees with that already observed for the BSSs of NGC~6397 (L12) 
and it points toward the occurrence of particle transport mechanisms driven by the radiative levitation. 
The enhancement of iron-peak elements due to this mechanism has been observed in HB stars 
hotter than $\sim$11000 K in Galactic GCs  \citep[see e.g.][]{Behr00a,Behr00b,Pace06}, and it is
typical of stars with shallow or no convective envelopes (as Population I MS stars, \citealt{Gebran2008, GebranMonier2008, Gebran2010}). 
{\sl M30 is the second cluster where the effects of radiative levitation are observed among the BSSs, confirming 
the results obtained for NGC~6397.}\\ 
(2)~The occurence of radiative levitation alters the surface chemical composition and prevents us to 
study the original chemical composition of BSSs hotter than $\sim$7800-8000 K. However, the 6 coldest BSSs 
in the sample have [Fe/H] ratios fully compatible with that of the cluster. This confirms that the surface 
chemical composition of these stars is not altered and allows us 
to discuss the chemical composition of these BSSs in terms of formation mechanisms. 
The most intriguing finding concerns the O abundance that in 4 out of 5 red cold BSSs results to be systematically 
lower than the typical values observed in the RGB and HB stars of the cluster.
The upper limits of [O/Fe] for those 4 red BSSs range from $-$0.13 dex down to $-$0.36 dex. 
and they are incompatible with the O abundances derived for the HB stars.
Unfortunately, the only cold BSS along the blue sequence is \#13000594, for which we are not able to derive significative O upper limits from 
the XSHOOTER spectra. 

\begin{figure}
\plotone{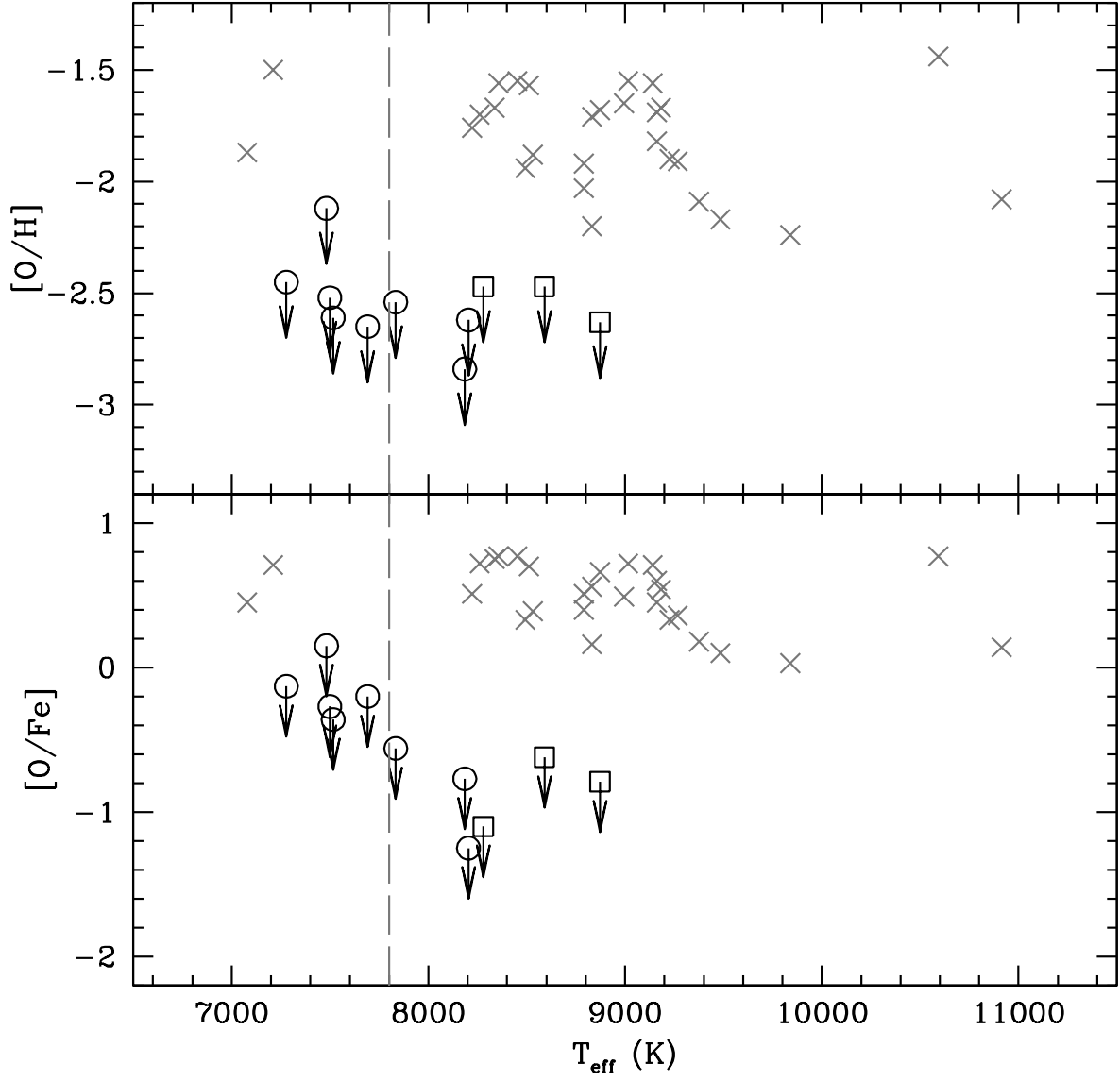}
\caption{O abundances for BSSs in the two sequences compared to those of HB stars (same symbols as in Fig.\ref{fem30}).} 
\label{oxm30}
\end{figure}

The very low O abundance in 4 red BSSs is incompatible with the 
O depletion associated with the second generation stars in this cluster (in fact, it is significantly
lower than that measured in HB stars and also in RGB stars; see Fig. \ref{oxm30} and \citealt{Carretta2009c}). This clearly suggests that other 
mechanisms able to reduce the O abundance occurred. 
In principle, a possibility could be that of the gravitational settling, which
is due to temperature, pressure and abundance gradients, leading to the sinking of the atoms.
Up to the present, no theoretical models taking into account the balance between radiative levitation
and gravitation settling exist in the literature for BSSs, so that it is not possible to predict the variation
of surface chemical abundances due to the occurrence of the two processes. On the base of a rough reasoning,
it could be possible that the gravitational settling leads to a simultaneous depletion of both O and Fe
and that a larger effect might be observed on Fe abundances because the Fe atoms are heavier than those of O.
In this case we would be able to discard that our results are the effect of the gravitational settling,
due to the fact that the Fe abundance for the O-depleted BSSs are fully compatible with those of normal cluster 
stars. However, even if we are not able to completely discard the possibility that the observed depletion is (at 
least partially) produced by the effect of gravitational settling, at the moment the hypothesis of a MT 
origin is the most likely, particularly according to the interpretation of the red sequence proposed by F09.
Unfortunately, the low metallicity of M30 does not allow us to measure C abundances, while 
the blue BSSs that should not be affected by radiative levitation are too faint to be properly 
observed with FLAMES. Thus, the present sample does not allow us to fully investigate whether the two BSS sequences 
of M30 harbor stars formed from the two different mechanisms,
but the strong O-depletion observed in 4 red BSSs is a relevant clue in support to this scenario.

\acknowledgements
{This research is part of the project COSMIC-LAB (www.cosmic-lab.eu) 
funded by the European Research Council (under contract ERC-2010-AdG-267675).
L. Lovisi wishes to thank ESO for the hospitality at ESO-Santiago provided during the preparation of the paper.
We thank the anonymous referee for his/her valuable suggestions.}

\begin{sidewaystable}
\scriptsize
\centering
\medskip
\rotatebox{90}{}
\setlength\tabcolsep{4pt}
\begin{tabular}{cccccccccccccc}
\hline
\hline
ID & RA & DEC & V & I & T$_{eff}$ & $\log(g)$ & RV & v$_{e}$ $\sin(i)$ & [Fe/H] & [Mg/H] & [O/H] & [Ti/H] & Notes\\ 
    & (degrees) & (degrees) & & & (K)&  & ($\kms$) & ($\kms$) & & & &\\
\hline
11000416R &   325.0948074 &  -23.1770660 &   17.92 &  17.54 &    7482 & 4.2 &    -186.0$\pm$1.7 &  5$\pm$1 &  -2.28 $\pm$0.07*&  -1.90$\pm$0.10 &  $<$-2.12 &  -2.10$\pm$0.09&\\
11000978R &   325.0910212 &  -23.1764598 &   16.95 &  16.71 &    8204 & 4.0 &    -183.4$\pm$3.3 &  25$\pm$2&  -1.37 $\pm$0.07 &  -2.30$\pm$0.10 &  $<$-2.62 &  -1.14$\pm$0.11&\\
11001895R &   325.0931321 &  -23.1791741 &   16.94 &  16.70 &    8185 & 4.0 &    -184.5$\pm$0.8 &  9$\pm$1 &  -2.07 $\pm$0.09 &  -2.23$\pm$0.10 &  $<$-2.84 &  -1.52$\pm$0.06&\\
11002072B &   325.0910559 &  -23.1783920 &   18.01 &  17.72 &    7980 & 4.4 &    -180.0$\pm$1.0 &   $<$30  &  -1.80 $\pm$0.14 &  -1.73$\pm$0.14 &         - &  -             & XSH\\
11002171R &   325.0913880 &  -23.1786920 &   17.72 &  17.38 &    7516 & 4.1 &    -184.7$\pm$1.0 &  5$\pm$2 &  -2.25 $\pm$0.10 &  -2.30$\pm$0.10 &  $<$-2.61 &  -1.85$\pm$0.06&\\
11003263R &   325.0932156 &  -23.1811967 &   17.61 &  17.30 &    7691 & 4.1 &    -189.0$\pm$1.2 &  7$\pm$3 &  -2.45 $\pm$0.07 &  -2.50$\pm$0.10 &  $<$-2.65 &  -2.06$\pm$0.07&\\
11004551R &   325.0903930 &  -23.1821092 &   17.65 &  17.28 &    7499 & 4.1 &    -180.3$\pm$0.8 &  4$\pm$1 &  -2.25 $\pm$0.14 &  -2.12$\pm$0.10 &  $<$-2.52 &  -1.86$\pm$0.07&\\
12000384R &   325.0987846 &  -23.1772652 &   17.32 &  17.06 &    7834 & 4.1 &    -183.0$\pm$1.1 &  0$\pm$1 &  -1.98 $\pm$0.20 &  -1.83$\pm$0.10 &  $<$-2.54 &  -1.77$\pm$0.09&\\
12001595R &   325.0957475 &  -23.1753040 &   17.25 &  16.86 &    7278 & 3.9 &    -183.1$\pm$1.6 &  5$\pm$1 &  -2.32 $\pm$0.06 &  -1.90$\pm$0.10 &  $<$-2.45 &  -2.27$\pm$0.06&\\
12002629B &   325.0938526 &  -23.1725247 &   17.41 &  17.29 &    8872 & 4.4 &    -184.0$\pm$0.7 &  10$\pm$1&  -1.84 $\pm$0.11 &  -1.81$\pm$0.10 &  $<$-2.63 &  -1.58$\pm$0.08&\\
12005407B &   325.0911142 &  -23.1590309 &   17.68 &  17.46 &    8204 & 4.3 &       -           &   $>$90  &             -    &        -        &         - &    -           & W Uma\\
12006406B &   325.0871233 &  -23.1573428 &   18.10 &  17.78 &    7889 & 4.4 &    -188.8$\pm$1.0 &   $<$30  &  -1.70$\pm$0.14  &  -1.63$\pm$0.14 &         - &   -            &XSH\\
13000594B &   325.1108234 &  -23.1814669 &   18.26 &  17.90 &    7691 & 4.4 &    -182.5$\pm$1.0 &   $<$30  &    $<$-2.10      &   $<$-2.30      &         - &   -            &XSH\\
14000585B &   325.1007172 &  -23.1796372 &   17.77 &  17.56 &    8279 & 4.4 &    -185.7$\pm$1.4 &  15$\pm$2&  -1.37 $\pm$0.10 &  -2.10$\pm$0.10 &  $<$-2.47 &  -1.12$\pm$0.07&\\
14002580B &   325.0993229 &  -23.1927703 &   17.45 &  17.31 &    8590 & 4.3 &    -186.5$\pm$0.4 &  23$\pm$4&  -1.85 $\pm$0.06 &  -1.25$\pm$0.10 &  $<$-2.47 &  -1.62$\pm$0.11&\\
\hline
\end{tabular}
\caption{Identification numbers (with R and B indicating BSSs in the red and blue sequence respectively), coordinates, magnitudes, effective temperatures, 
surface gravities, radial and rotational velocities, Fe, Mg, O and Ti abundances of the BSS sample. In the last column XSH labels BSSs observed 
with XSHOOTER. For BSS \#11000416, no FeII lines have been detected, thus we attributed to this star the [Fe/H] inferred from the RGB stars.}
\label{bssm30} 
\end{sidewaystable} 

\begin{sidewaystable}
\scriptsize
\centering
\medskip
\rotatebox{90}{}
\setlength\tabcolsep{5pt}
\begin{tabular}{ccccccccccccc}
\hline
\hline
ID & RA & DEC & V & I & T$_{eff}$ & $\log(g)$ & v$_{t}$ & RV & v$_{e}$ $\sin(i)$ & [Fe/H] & [Mg/H] & [Ti/H]\\ 
   & (degrees) & (degrees) & & & (K)& &($\kms$)& ($\kms$) & ($\kms$) &&&\\
\hline
10201736&    325.0986591& -23.1635326&    15.31&  14.38&   5117& 2.3& 1.6&   -177.0$\pm$0.3& 0$\pm$1 & -2.29$\pm$0.21&    -2.01$\pm$0.18&   -2.00$\pm$0.16\\
10202312&    325.0990668& -23.1572188&    15.47&  14.52&   5140& 2.4& 1.6&   -183.4$\pm$0.4& 0$\pm$1 & -2.39$\pm$0.20&    -1.93$\pm$0.18&   -1.98$\pm$0.18\\
10300934&    325.1168324& -23.1802542&    15.69&  14.77&   5176& 2.5& 1.6&   -178.3$\pm$0.4& 0$\pm$1 & -2.34$\pm$0.21&    -1.84$\pm$0.18&   -2.09$\pm$0.18\\
10301439&    325.1149684& -23.1749077&    14.97&  13.99&   5047& 2.2& 1.6&   -188.4$\pm$0.4& 0$\pm$1 & -2.28$\pm$0.21&    -1.82$\pm$0.18&   -1.94$\pm$0.18\\
10400043&    325.0907822& -23.1937077&    15.79&  14.91&   5200& 2.6& 1.5&   -188.4$\pm$0.4& 0$\pm$1 & -2.32$\pm$0.22&    -1.71$\pm$0.18&   -2.07$\pm$0.17\\
12001703&    325.1028333& -23.1626725&    16.67&  15.82&   5333& 3.0& 1.4&   -179.1$\pm$0.4& 6$\pm$1 & -2.36$\pm$0.21&    -1.99$\pm$0.18&   -2.13$\pm$0.18\\
13002556&    325.1197681& -23.1771157&    16.43&  15.57&   5297& 2.9& 1.5&   -185.3$\pm$0.5& 0$\pm$1 & -2.36$\pm$0.21&    -2.07$\pm$0.18&   -1.90$\pm$0.17\\
13003187&    325.1186601& -23.1727381&    16.91&  16.05&   5370& 3.0& 1.4&   -184.1$\pm$0.7& 0$\pm$1 & -2.30$\pm$0.20&    -1.93$\pm$0.18&   -2.15$\pm$0.16\\
13004219&    325.1264165& -23.1688178&    16.21&  15.32&   5272& 2.8& 1.5&   -184.6$\pm$0.4& 0$\pm$1 & -2.22$\pm$0.21&    -2.07$\pm$0.18&   -2.08$\pm$0.17\\
14000326&    325.0915862& -23.1934773&    16.94&  16.09&   5370& 3.1& 1.4&   -192.3$\pm$0.4& 0$\pm$1 & -2.43$\pm$0.22&    -1.99$\pm$0.18&   -1.91$\pm$0.19\\
\hline
\end{tabular}
\caption{Identification numbers, coordinates, magnitudes, effective temperatures, surface gravities, microturbulent velocities, radial and rotational 
velocities, Fe, Mg and Ti abundances of the RGB sample. This table is available in its entirety in a machine-readable form in the online journal. 
A portion is shown here for guidance regarding its form and content.}
\label{rgbm30} 
\end{sidewaystable}

\begin{sidewaystable}
\scriptsize
\centering
\medskip
\rotatebox{90}{}
\setlength\tabcolsep{5pt}
\begin{tabular}{ccccccccccccc}
\hline
\hline
ID & RA & DEC & V & I & T$_{eff}$ & $\log(g)$ & RV & v$_{e}$ $\sin(i)$ & [Fe/H] & [Mg/H] & [O/H] &[Ti/H]\\ 
   & (degrees) & (degrees) & & &(K) &($\kms$)& ($\kms$) & ($\kms$) &&&\\
\hline
10201925 &  325.0989565 &   -23.1613142 & 	16.05 & 	 16.03 & 	10914 &  3.7 &   -191.8$\pm$  1.0 &    0$\pm$1&  -              &    -2.10$\pm$0.05  & -2.08$\pm$0.14 &  -     \\
10202614 &  325.0891261 &   -23.1711663 & 	15.82 & 	 15.75 & 	10069 &  3.6 &   -183.9$\pm$  2.1 &   47$\pm$4&  -      	&    -2.04$\pm$0.05  &  -  &  -   \\  
10203922 &  325.0904138 &   -23.1512884 & 	15.84 & 	 15.82 & 	10186 &  3.6 &   -186.9$\pm$  1.3 &   40$\pm$1&  -     		&    -1.94$\pm$0.06  &  -  &  -   \\
10301333 &  325.1135774 &   -23.1751181 & 	15.80 & 	 15.76 & 	10023 &  3.6 &   -192.8$\pm$  1.4 &   36$\pm$1&  -     		&    -2.03$\pm$0.06  &  -  &  -   \\
10301793 &  325.1201773 &   -23.1736029 & 	15.62 & 	 15.57 & 	 9376 &  3.4 &   -177.9$\pm$  0.6 &   19$\pm$1&  -2.22$\pm$0.05 &    -2.01$\pm$0.05  & -2.09$\pm$0.03 &  -     \\
10400383 &  325.0940697 &   -23.1903654 & 	15.32 & 	 15.16 & 	 8222 &  3.2 &   -183.2$\pm$  0.7 &   29$\pm$2&  -     		&    -1.83$\pm$0.08  & -1.76$\pm$0.20 &  -     \\
10400762 &  325.0974077 &   -23.1873751 & 	15.57 & 	 15.51 & 	 9226 &  3.4 &   -186.4$\pm$  0.4 &    0$\pm$1&  -2.23$\pm$0.05 &    -1.96$\pm$0.05  & -1.90$\pm$0.15 &  -1.93$\pm$0.14 \\
10401890 &  325.1056998 &   -23.1838967 & 	15.77 & 	 15.70 & 	 9931 &  3.6 &   -190.6$\pm$  1.2 &   28$\pm$1&  -     		&    -2.09$\pm$0.07  &  - &  -    \\
10402581 &  325.1033078 &   -23.1968969 & 	15.18 & 	 14.84 & 	 7379 &  3.0 &   -180.3$\pm$  1.2 &   35$\pm$1&  -     		&    -1.74$\pm$0.05  &  - &  -    \\
20100075 &  325.0861829 &   -23.2052191 & 	15.33 & 	 15.16 & 	 8241 &  3.2 &   -181.8$\pm$  0.7 &   35$\pm$2&  -     		&    -2.01$\pm$0.09  &  - &  -    \\
\hline
\end{tabular}
\caption{Identification numbers, coordinates, magnitudes, effective temperatures, surface gravities, radial and rotational velocities, Fe, Mg, O and Ti 
abundances of the HB sample. This table is available in its entirety in a machine-readable form in the online journal. 
A portion is shown here for guidance regarding its form and content.}
\label{hbm30} 
\end{sidewaystable} 

\begin{sidewaystable}
\centering
\medskip
\rotatebox{90}{}
\setlength\tabcolsep{5pt}
\begin{tabular}{cccc}
\hline
\hline
wavelength  & ion & log(gf) & E.P.\\
   \AA     &      &         & eV \\
\hline
4395.031  & Ti~II  &  -0.540  &  1.08 \\    
4399.765  & Ti~II  &  -1.190  &  1.24 \\    
4417.714  & Ti~II  &  -1.190  &  1.16 \\    
4443.801  & Ti~II  &  -0.720  &  1.08 \\    
4450.482  & Ti~II  &  -1.520  &  1.08 \\    
4468.492  & Ti~II  &  -0.600  &  1.13 \\    
4481.126  & Mg~II  &   0.749  &  8.86 \\    
4491.405  & Fe~II  &  -2.640  &  2.86 \\    
4501.270  & Ti~II  &  -0.770  &  1.12 \\    
4508.288  & Fe~II  &  -2.350  &  2.86 \\    
4515.339  & Fe~II  &  -2.360  &  2.84 \\    
4520.224  & Fe~II  &  -2.620  &  2.81 \\    
4522.634  & Fe~II  &  -1.990  &  2.84 \\    
4533.960  & Ti~II  &  -0.530  &  1.24 \\    
4549.621  & Ti~II  &  -0.110  &  1.58 \\    
4555.893  & Fe~II  &  -2.250  &  2.83 \\    
4563.757  & Ti~II  &  -0.690  &  1.22 \\    
4571.971  & Ti~II  &  -0.320  &  1.57 \\    
4583.837  & Fe~II  &  -1.740  &  2.81 \\    
4589.947  & Ti~II  &  -1.790  &  1.24 \\ 	 
4923.927  & Fe~II  &  -1.504  &  2.89 \\
5018.440  & Fe~II  &  -1.345  &  2.89 \\ 
7771.954  &   O~I  &   0.369  &  9.15 \\     
7774.176  &   O~I  &   0.223  &  9.15 \\     
7775.398  &   O~I  &   0.001  &  9.15 \\     
\hline
\end{tabular}
\caption{Wavelength, atomic number, logarithm of the oscillator strength and excitation potential of the main lines
used to derive chemical abundances for the BSS sample. Mg~II line at 4481.126 \AA\ and Fe~II lines at 4583.837, 4923.927 and 5018.440
\AA\ have been used to derive abundances (or upper limits) also for the XSOOTER targets.}
\label{bsslines} 
\end{sidewaystable}

\begin{sidewaystable}
\centering
\medskip
\small
\rotatebox{90}{}
\setlength\tabcolsep{5pt}
\begin{tabular}{cccc}
\hline
\hline
wavelength  & ion & log(gf) & E.P.\\
   \AA     &               &         &  eV\\
\hline
4383.545   &   Fe~I  &  0.200	& 1.48\\    
4389.245   &   Fe~I  & -4.583	& 0.05\\    
4394.059   &  Ti~II  & -1.780	& 1.22\\    
4395.839   &  Ti~II  & -1.930	& 1.24\\    
4399.765   &  Ti~II  & -1.190	& 1.24\\    
4411.929   &  Ti~II  & -2.520	& 1.22\\    
4415.122   &   Fe~I  & -0.615	& 1.61\\    
4417.714   &  Ti~II  & -1.190	& 1.16\\    
4418.331   &  Ti~II  & -1.970	& 1.24\\    
4433.219   &   Fe~I  & -0.730	& 3.65\\    
4443.801   &  Ti~II  & -0.720	& 1.08\\    
4447.717   &   Fe~I  & -1.342	& 2.22\\    
4450.482   &  Ti~II  & -1.520	& 1.08\\    
4468.492   &  Ti~II  & -0.600	& 1.13\\    
4484.220   &   Fe~I  & -0.864	& 3.60\\    
4489.739   &   Fe~I  & -3.966	& 0.12\\    
4493.523   &  Ti~II  & -3.020	& 1.08\\     
4494.563   &   Fe~I  & -1.136	& 2.20\\    
4501.270   &  Ti~II  & -0.770	& 1.12\\    
4518.332   &  Ti~II  & -2.910	& 1.08\\    
4524.679   &  Ti~II  & -2.690	& 1.23\\    
4529.480   &  Ti~II  & -1.640	& 1.57\\    
4531.148   &   Fe~I  & -2.155	& 1.48\\    
4545.134   &  Ti~II  & -2.980	& 1.13\\    
4547.847   &   Fe~I  & -1.010	& 3.55\\    
4554.031   &  Ba~II  &  0.163   & 0.00\\
4563.757   &  Ti~II  & -0.690	& 1.22\\    
4571.971   &  Ti~II  & -0.320	& 1.57\\    
4571.096   &   Mg~I  & -5.691	& 0.00\\    
\hline
\end{tabular}
\caption{Wavelength, atomic number, logarithm of the oscillator strength and excitation potential of the main lines
used to derive chemical abundances for the RGB sample.}
\label{rgblines} 
\end{sidewaystable}

\begin{sidewaystable}
\centering
\medskip
\rotatebox{90}{}
\setlength\tabcolsep{5pt}
\begin{tabular}{cccc}
\hline
\hline
wavelength  & ion & log(gf) & E.P.\\
   \AA     &               &         &  eV\\
\hline
4385.387 &  Fe~II  &  -2.580  &  2.78\\    
4395.031 &  Ti~II  &  -0.540  &  1.08\\     
4399.765 &  Ti~II  &  -1.190  &  1.24\\    
4416.830 &  Fe~II  &  -2.600  &  2.78\\    
4417.714 &  Ti~II  &  -1.190  &  1.16\\    
4443.801 &  Ti~II  &  -0.720  &  1.08\\       
4468.492 &  Ti~II  &  -0.600  &  1.13\\    
4481.126 &  Mg~II  &   0.749  &  8.86\\    
4489.183 &  Fe~II  &  -2.970  &  2.83\\    
4491.405 &  Fe~II  &  -2.640  &  2.86\\    
4501.270 &  Ti~II  &  -0.770  &  1.12\\    
4508.288 &  Fe~II  &  -2.350  &  2.86\\    
4515.339 &  Fe~II  &  -2.360  &  2.84\\    
4520.224 &  Fe~II  &  -2.620  &  2.81\\    
4522.634 &  Fe~II  &  -1.990  &  2.84\\     
4533.960 &  Ti~II  &  -0.530  &  1.24\\    
4541.524 &  Fe~II  &  -2.970  &  2.86\\    
4555.893 &  Fe~II  &  -2.250  &  2.83\\     
4563.757 &  Ti~II  &  -0.690  &  1.22\\    
4571.971 &  Ti~II  &  -0.320  &  1.57\\    
4576.340 &  Fe~II  &  -2.920  &  2.84\\    
4582.835 &  Fe~II  &  -3.062  &  2.84\\    
4583.837 &  Fe~II  &  -1.740  &  2.81\\     
7771.954 &    O~I  &   0.369  & 9.15\\     
7774.176 &    O~I  &   0.223  & 9.15\\     
7775.398 &    O~I  &   0.001  & 9.15\\     
\hline
\end{tabular}
\caption{Wavelength, atomic number, logarithm of the oscillator strength and excitation potential of the main lines
used to derive chemical abundances for the HB sample.}
\label{hblines} 
\end{sidewaystable}

\end{document}